\documentclass{ws-procs961x669} 
\usepackage{dirtytalk}

\begin{document}
\title{A new perspective on cosmology through Supernovae Ia and Gamma Ray Bursts}

\author{B. De Simone}

\address{Department of Phyisics "E. R. Caianiello", University of Salerno,\\
Via Giovanni Paolo II, 132, 84084, Fisciano, Salerno, Italy\\
E-mail: b.desimone1@studenti.unisa.it\\
www.df.unisa.it}

\author{V. Nielson}

\address{SLAC National Accelerator Laboratory, 2575 Sand Hill Road, Menlo Park, CA 94025, USA\\ 
Astronomy Department, University of Michigan, Ann Arbor, MI 48109, USA}

\author{E. Rinaldi}

\address{Physics Department, University of Michigan, Ann Arbor, MI 48109, USA\\
RIKEN Cluster for Pioneering Research, Theoretical Quantum Physics Laboratory, Wako, Saitama 351-0198, Japan\\
RIKEN iTHEMS, Wako, Saitama 351-0198, Japan\\
RIKEN Center for Quantum Computing, Wako, Saitama 351-0198, Japan}

\author{M. G. Dainotti}

\address{National Astronomical Observatory of Japan, 2 Chome-21-1 Osawa, Mitaka, Tokyo 181-8588, Japan\\
The Graduate University for Advanced Studies, SOKENDAI, Shonankokusaimura, Hayama, Miura District, Kanagawa 240-0193, Japan\\
Space Science Institute, Boulder, CO, USA\\
\footnote{Corresponding author}E-mail: maria.dainotti@nao.ac.jp\\
www.guas-astronomy.jp/eng}

\begin{abstract}
The actual knowledge of the structure and future evolution of our universe is based on the use of cosmological models, which can be tested through the so-called 'probes', namely astrophysical phenomena, objects or structures with peculiar properties that can help to discriminate among different cosmological models. Among all the existing probes, of particular importance are the Supernovae Ia (SNe Ia) and the Gamma Ray Bursts (GRBs): the former are considered among the best standard candles so far discovered but suffer from the fact that can be observed until redshift $z=2.26$, while the latter are promising standardizable candles which have been observed up to $z=9.4$, surpassing even the farthest quasar known to date, which is at $z=7.64$. The standard candles can be used to test the cosmological models and to give the expected values of cosmological parameters, in particular the Hubble constant value. The Hubble constant is affected by the so-called \say{Hubble constant tension}, a discrepancy in more than 4 $\sigma$ between its value measured with local probes and its value measured through the cosmological probes. The increase in the number of observed SNe Ia, as well as the future standardization of GRBs through their correlations, will surely be of help in alleviating the Hubble constant tension and in explaining the structure of the universe at higher redshifts. A promising class of GRBs for future standardization is represented by the GRBs associated with Supernovae Ib/c, since these present features similar to the SNe Ia class and obey a tight correlation between their luminosity at the end of the plateau emission in X-rays and the time at the end of the plateau in the rest-frame.   
\end{abstract}

\keywords{Gamma Ray Bursts, Supernovae Ia, Cosmology, Correlations, Hubble constant, Hubble tension}

\bodymatter

\section*{Introduction}
Modern cosmology is based on the so-called $\Lambda$CDM model: this is the standard cosmological model based on the Cold Dark Matter (CDM) and containing the cosmological constant $\Lambda$ that describes the dark energy contribution to the expansion of the cosmos. This model has been widely accepted in the scientific community and can predict the accelerated expansion of the universe, as proved by the outstanding works of Riess et al. (1998) \cite{Riess1998} and Perlmutter et al. (1999) \cite{Perlmutter1999} where this effect was demonstrated through the use of SNe Ia. Despite being well received, the $\Lambda$CDM suffers from some open problems, in particular the Hubble constant ($H_0$) tension. This is the discrepancy, in more than 4 $\sigma$, between the value of $H_0$ measured with local probes (Cepheids and SNe Ia) and the value obtained through the Cosmic Microwave Background (CMB) radiation data measured by the Planck satellite. To solve this issue, many approaches have been proposed (alternative cosmological models, refined measurements of $H_0$, etc.), but it is necessary to use reliable probes for testing the cosmological models. From this perspective, the so-called standard candles are very important: these are astrophysical objects or events which have a fixed luminosity or that obey an intrinsic relation between the luminosity and some of the other parameters that do not depend on the luminosity themselves. To date, different objects have been standardized, in particular the SNe Ia which have a nearly uniform peak luminosity, but the problem is that these can cover only a relatively small range of redshifts: the farthest SN Ia so far observed has a redshift of $z=2.26$ \cite{Rodney2016}. The GRBs may have a key role in the future development of cosmology since they have been observed at higher redshifts than SNe Ia and quasars (currently the farthest quasar being at redshift 7.64 \cite{Wang2021}) and are able to further extend the Hubble diagram (the distance moduli versus the redshift). A particular class of GRBs, the GRBs with a plateau emission (namely, a relatively flat section of the lightcurve that follows the prompt emission and precedes the afterglow, observed not only in X-rays and optical but also in $\gamma$-rays\cite{DainottiOmodei2021}), has proven to be a promising standardizable candle through the application of tight correlations among the GRB lightcurve properties, such as the intrinsic and unbiased correlation between the luminosity at the end of the plateau emission in the X-rays, $L_X$, with the time at the end of the plateau emission in the rest-frame, $T^*_X$\cite{Dainotti2008,Dainotti2010,Dainotti2011a,Dainotti2013a,Dainotti2013b,Dainotti2015a,DelVecchio2016,Dainotti2017a} (which was later confirmed also in the optical\cite{Dainotti2020b}) and the unbiased one between $L_X$ and the 1s peak prompt luminosity, $L_{peak}$\cite{Dainotti2011b,Dainotti2015b}. A combination of these two correlations led to the so-called 3D fundamental plane relation, namely a tight correspondence between $L_X$, $T^*_X$, and $L_{peak}$\cite{Dainotti2016,Dainotti2017c,Dainotti2020a}. For a review of these correlations, see Dainotti \& Del Vecchio (2017)\cite{DainottiDelVecchio2017}, Dainotti et al. (2018a,2018b)\cite{Dainotti2018a,Dainotti2018b}, and Dainotti (2019)\cite{Dainotti2019book}. The GRB correlations are not only useful as cosmological tools, but also as discriminant among theoretical models to explain the GRBs origin, emission mechanism, and the nature of their environment \cite{Kumar2008,Rowlinson2014,Rea2015,Srinivasaragavan2020,Dainotti2021closure,DainottiPetrosianBowden2021}. 
In particular, the plateau emission and the correlations that involve its properties strongly suggest how the typical magnetic fields and spin periods of the magnetars (namely, a fast rotating neutron star) could explain the plateau emission itself (Rowlinson et al. 2014)\cite{Rowlinson2014}. In a later paper, Rea et al. (2015)\cite{Rea2015} show that the magnetar model can be reconciled within the GRB emission in the plateau only if supermagnetar with high magnetic field strengh are allowed. In a successive paper of Stratta et al. (2018)\cite{Stratta2018} for the first time a non-ideal modelling of spindown magnetar is fitted to the afterglow data with a statistical sample of 40 Long GRBs (LGRBs) with a well-defined plateau and 13 Short GRBs (SGRBs) including the short with extended emission. The conclusion reached in that paper is that SGRBs including the SGRBs with Extended Emission (SEE) and LGRBs can be explained within the magnetar model but with the difference that the LGRBs occupy a lower end in the magnetic field-spin period plane compared to the SGRBs which present a higher spin, P, and a higher magnetic field, B. The correlation between magnetic field and spin period follow the established physics of the spin-up line for accreting neutron star in Galactic binary systems. The B-P relation obtained perfectly matches spin-up line predictions for the magnetar model with mass accretion rates expected in the GRB prompt phase. The latter are $\sim 11-14$ orders of magnitude higher
than those inferred for the Galatic accreting NSs. Thus, correlations are useful to cast more light on the physics of GRBs and this represents an effective support towards the future standardization of these transient phenomena. Before introducing the contribution of this work to the age-old issue of the $H_0$ tension, it is important to summarize the current state of the research carried on to alleviate it and, in the future, to solve it. There is a wide range of possibilities behind the observed discrepancy in the measured values of $H_0$. Many authors have tested cosmological and astrophysical models which represent a deviation from the $\Lambda$CDM or the standard knowledge of the elementary particles and interactions in the universe \cite{Li2015,Tomita2017,Tomita2018,Wang2018,Tomita2019,Tomita2020,Vagnozzi2020,Adil2021,Aghababaei2021,Aghaei2021,Alestas2021a,Alestas2021c,Bag2021,Bansal2021,Benisty2021,Blinov2021,Cedeno2021a,Cedeno2021b,Cuesta2021,DiBari2021,Fanizza2021,Fung2021,Geng2021,Ghosh2021,Gurzadyan2021,Hansen2021,HernandezAlmada2021,Jiang2021,Joseph2021,Krishnan2021b,Li2021,Liu2021a,Martin2021,Najera2021,Nguyen2020,Nojiri2021,Odintsov2021,Palle2021,Parnovsky2021b,Petronikolou2021,Reyes2021,Shokri2021,Sola2021,Tian2021,Vagnozzi2021,Wang2021,Yang2021,Ye2021a,Ye2021b,Zhang2021}. The tension may be alleviated also through the introduction of new cosmological probes and merged data from different sources which led to more precise constraints on the $H_0$ value \cite{Krishnan2020,Alestas2021b,Asencio2021,BohuaLi2021,Castello2021,Ferree2021,FranchinoVinas2021,Freedman2021,Galli2021,Gerardi2021,Haslbauer2021,Hou2021,Javanmardi2021,Krishnan2021a,Liu2021b,Mehrabi2021b,Normann2021,Perivolaropoulos2021,Peterson2021,Roth2021,RuizZapatero2021,Sharov2020,Thakur2021,Thiele2021,Wu2021,Zhao2021}, while other works focused on the issue of the astrophysical biases as a probable cause for the tension \cite{Parnovsky2021a,Baldwin2021}. Alternative approaches are based on the parametrization of cosmological observables \cite{Renzi2020,Cai2021,Dinda2021,EscamillaRivera2021,Krishnan2021c,Marra2021,Mehrabi2021a,Ren2021,Shrivastava2021}, the proposal of new representative models for the cosmological parameters \cite{Bernal2021}, and the use of machine learning techniques \cite{Huber2021}. For a complete review of the state-of-the-art of the $H_0$ tension, see Di Valentino et al. (2021)\cite{DiValentino2021} and Perivolaropoulos \& Skara (2021)\cite{PerivolaropoulosSkara2021}.\\ 
The current proceeding focuses on two main topics concerning cosmology with SNe Ia and GRBs: (1) the investigation of the $H_0$ tension through a binning approach on the Pantheon sample of SNe Ia \cite{DainottiDeSimone2021} and (2) the use of the GRB fundamental planes in optical and X-ray as cosmological distance indicators and the discussion of their future use as standardizable candles.

\section{On the Hubble constant tension in the SNe Ia Pantheon sample}
The $H_0$ tension is the discrepancy in more than 4 $\sigma$ between the value of $H_0$ estimated with local probes, such as SNe Ia and Cepheids ($H_0=74.03 \pm 1.42$ $km$ $s^{-1} Mpc^{-1}$), and the value of $H_0$ obtained with the Planck Cosmic Microwave Background (CMB) radiation ($H_0=67.4 \pm 0.5$ $km$ $s^{-1} Mpc^{-1}$). It represents one of the most important open problems in modern cosmology. To investigate it, we divided the Pantheon sample (Scolnic et al. 2018\cite{Scolnic2018}), a collection of 1048 spectroscopically confirmed SNe Ia, into 3, 4, 20, and 40 bins ordered in redshift. In our analysis, we considered two cosmological models: the standard $\Lambda$CDM model and the $w_0 w_a$CDM model, where the equation of state parameter is expressed according to the Chevallier-Polarski-Linder parametrization ($w(z)=w_0+w_a*z/(1+z)$)\cite{Linder2003,Chevallier2001}. 

\begin{figure}
    \centering
    \includegraphics[scale=0.14]{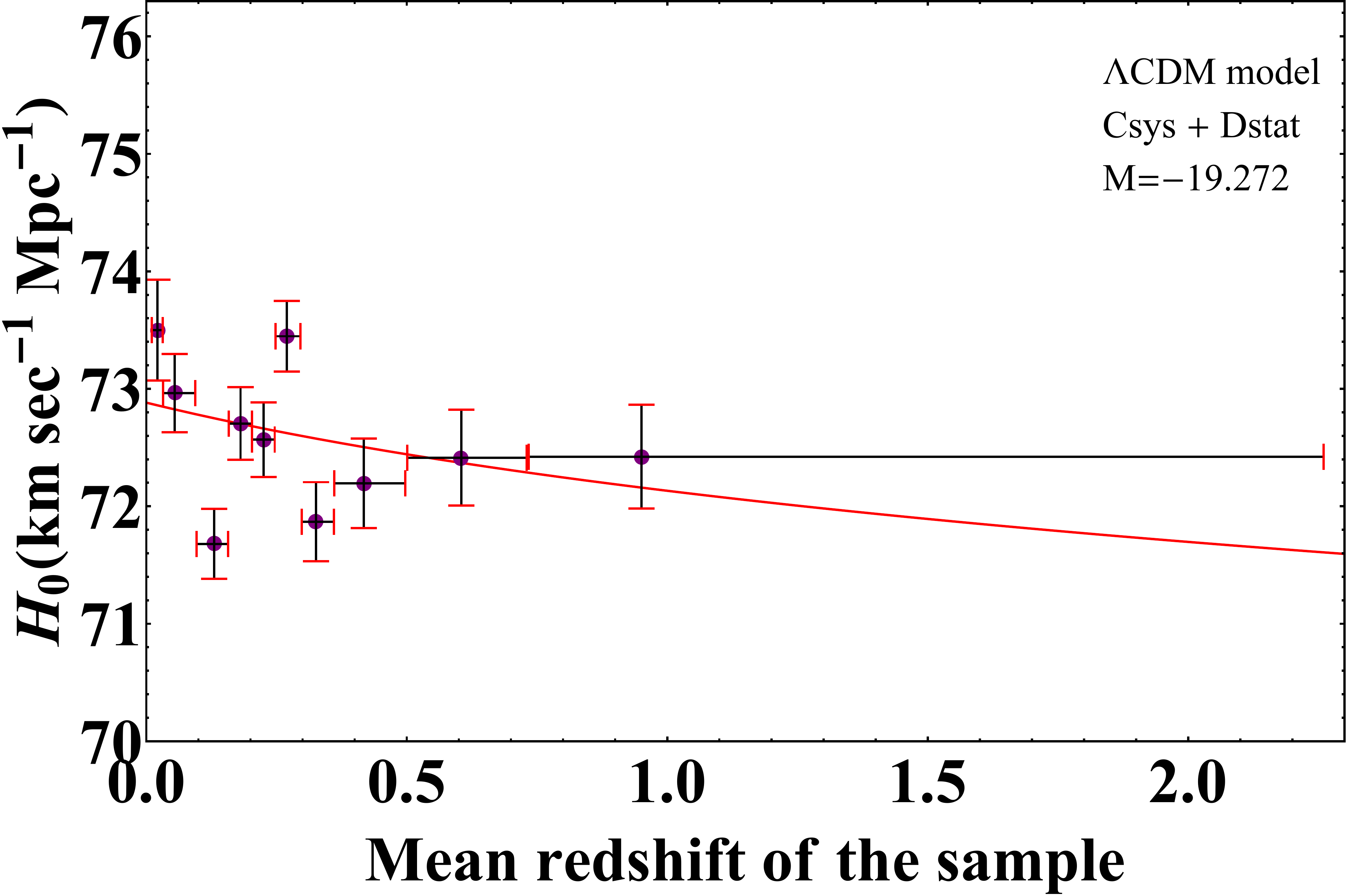}
    \includegraphics[scale=0.154]{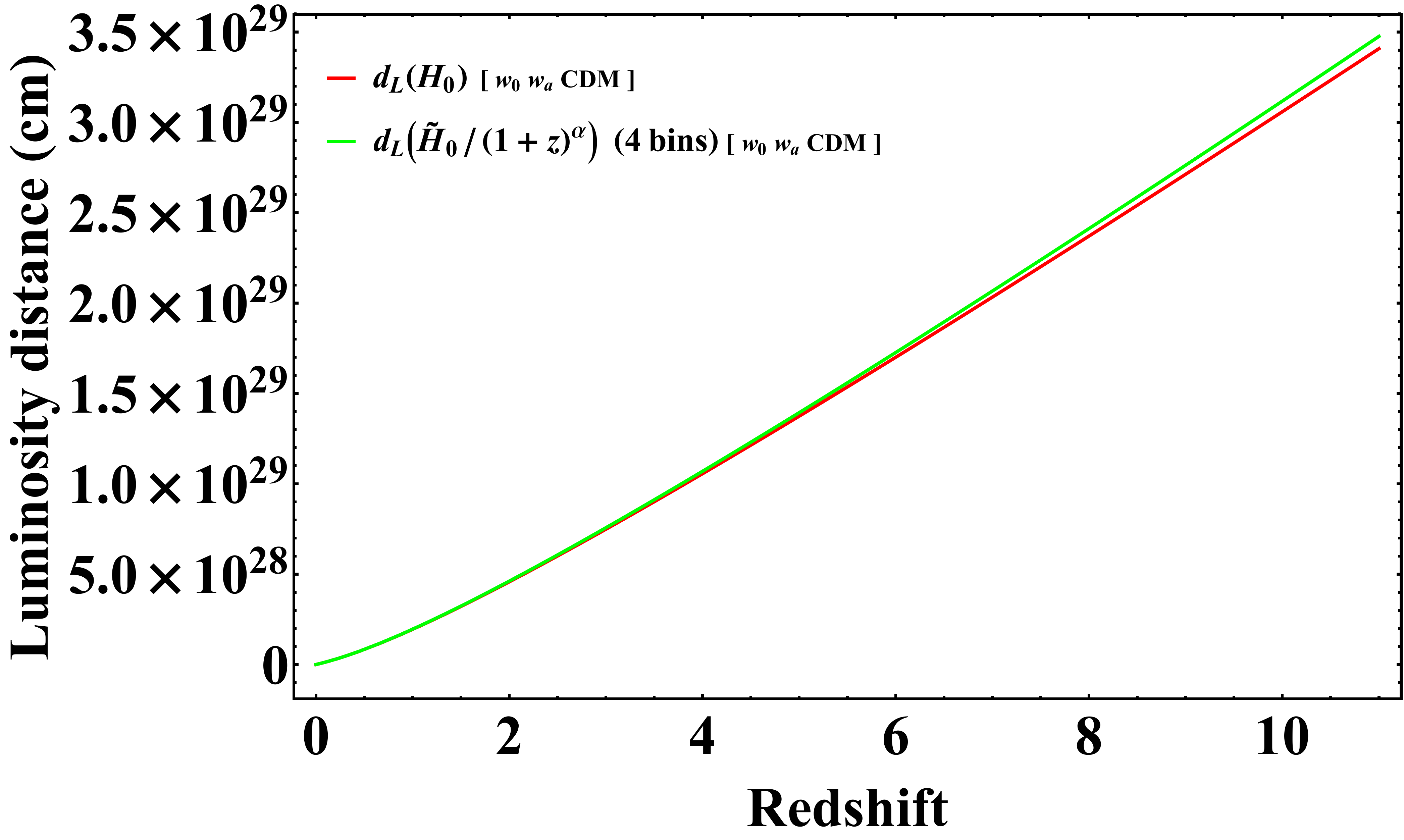}    
    \caption{\textbf{Left panel.} The fitting of the $H_0$ values with redshift in the case of 10 bins of SNe, assuming the $\Lambda$CDM model, using the full covariance matrix composed of systematic ($C_{sys}$) and statistical ($D_{stat}$) uncertainties\cite{Scolnic2018}, and fixing $M=-19.272$ such that $H_0=73.5 km$ $s^{-1}$ $Mpc^{-1}$ in the first bin. The slow decreasing trend is plain to see. \textbf{Right panel.} The comparison between the luminosity distance formula corrected with the $g(z)$ (in green) and the luminosity distance given by the standard $w_{0}w_{a}$CDM model (in red) as functions of the redshift.}
    \label{fig1}
\end{figure}

Due to the degeneracy between the $H_0$ and the fiducial absolute magnitude of SNe Ia, $M$, we set $M$ in such a way that locally the $H_0=73.5 km$ $s^{-1}$ $Mpc^{-1}$. The fiducial values of the total matter density parameters have been assumed as $\Omega_M=0.298$ (for the $\Lambda$CDM model) and $\Omega_M=0.308$ (for the $w_0 w_a$CDM model). Considering both the cosmological models, for each bin in the 3, 4, 20, and 40 divisions we performed a $\chi^2$ minimization followed by a Monte Carlo Markov Chain simulation to obtain the best-fit $H_0$ values together with their 1 $\sigma$ uncertainties. After obtaining these, we fitted them with the following functional form: $g(z)=H'_0/(1+z)^\alpha$, where $H’_0$ is the value of the Hubble constant at $z=0$ obtained with the fitting and $\alpha$ is the evolutionary parameter. We found that the $H_0$ in the Pantheon sample evolves slowly with the redshift, with an $\alpha$ coefficient in the order of $10^{-2}$ which is compatible with zero from 1.2 $\sigma$ to 2.0 $\sigma$. For example, in the left panel of Figure \ref{fig1} the case of fitting with 10 bins using the $\Lambda$CDM model is shown, while in the right panel a comparison between the standard luminosity distance in the $w_{0}w_{a}$CDM model and the corrected luminosity distance substituting $H_0$ with $g(z)$ is plotted.
Despite the $\alpha$ coefficients being compatible with zero in 3 $\sigma$, the highlighted decreasing trend may affect the cosmological results. Indeed, a modification of the luminosity distance formula where $H_0$ is replaced with the $g(z)$ form shows how the modified luminosity distance curve departs significantly from the $\Lambda$CDM canonical one at redshift $z\sim10$ (see the right panel of Figure \ref{fig1}). To check what would be the behavior of $H_0$ if the trend was real, we extrapolated its value from the fitting at the redshift of the most distant galaxy so far discovered ($z=11.09$, Oesch et al. 2016\cite{Oesch2016}) and at the redshift of the last scattering surface ($z=1100$). We obtained that the extrapolated value of $H_0$ at $z=1100$ is compatible in 1 $\sigma$ with the one obtained through the Planck CMB measurements. If we consider the discrepancy between the aforementioned values of $H_0$, namely $(74.03 \pm 1.42)$ and $(67.4 \pm 0.5)$ $km s^{-1} Mpc^{-1}$ and we compare it with the tension between the fitting values $H’_0$ (at $z=0$) and $H_0(z=1100)$, we find that our approach leads to alleviate the $H_0$ tension from 54\% to 72\% for the $\Lambda$CDM and the $w_0 w_a$CDM models, respectively. Thus, we not only found a way to alleviate the Hubble tension, but we provided also a plausible explanation for the observed discrepancy between the values of $H_0$ coming from probes at different redshifts. The observed decreasing trend may be due to hidden astrophysical biases, selection effects, or even the evolution of SNe Ia lightcurve parameters, as pointed out in Nicolas et al. (2021)\cite{Nicolas2021} where it was shown that the stretch parameter distribution of the Pantheon sample is affected by the drift with the redshift. If this is not the case, and the trend is real instead, the explanation for it may be found in the modified gravity theories. We propose that the $f(R)$ theories in the Jordan frame may explain the observed slow evolution of $H_0$. However, to prove if this trend is real or not, it is necessary to rely (a) on the future observations of SNe Ia through that will enrich the currently known samples of transient phenomena and (b) on the extension of the Hubble diagram up to redshift ranges that SNe Ia could not cover, thus calling for the use of high-z phenomena such as the GRBs. 

\section{Optical and X-ray GRB Fundamental Planes as Cosmological Distance Indicators}
GRBs, being observed at very high-z (up to $z=9.4$ so far\cite{Cucchiara2011}), have the potential to be employed as standardizable candles, extending the cosmological distance ladder beyond the redshift of SNe Ia. Therefore, we employ them to do so by first standardizing them using the 3D fundamental plane relation. 50 X-ray GRBs cut from a full sample of 222 define the platinum sample, and their corresponding fundamental plane parameter variables are determined by Monte Carlo Markov Chain sampling maintained by Gelman-Rubin statistical constraints. We use both the platinum sample and a well-defined sample of SNe Ia as a combination of probes to accurately constrain the matter content of the modern universe, $\Omega_M$. To do so, the plane parameters are allowed to vary simultaneously with $\Omega_M$ within the chain sampling. We find our best results with the addition of a third probe, namely, Baryon Acoustic Oscillations (BAOs), given their reliability as standard rulers. The combination of these three probes allows us to constrain $\Omega_M$ to $0.306 \pm 0.006$, assuming a $\Lambda$CDM cosmology. Furthermore, we test, for the first time in this research field, the novel 3D optical correlation as the extension of the 3D fundamental plane in X-ray wavelengths as a cosmological tool and check its applicability compared to that of the confirmed X-ray relation. In doing so, we find that our optical GRB sample is as efficacious in the determination of $\Omega_M$ as the platinum X-ray sample. To increase the precision on the estimate of $\Omega_M$, we consider redshift evolutionary effects to overcome common biases such as the Malmquist effect. We employ the reliable Efron \& Petrosian (1992)\cite{EfronPetrosian1992} statistical method to ensure that the correlation is intrinsic to the GRB physical mechanics and not due to selection bias. It is by accounting for this that we decrease the intrinsic scatter on the X-ray plane by 44.4\%, thus defining the tightest 3D GRB correlation involving plateau features in the literature to date. Using this corrected sample, we confirm a value of $0.306 \pm 0.006$ defines $\Omega_M$. 

\begin{figure}
    \centering
    \includegraphics[scale=0.18]{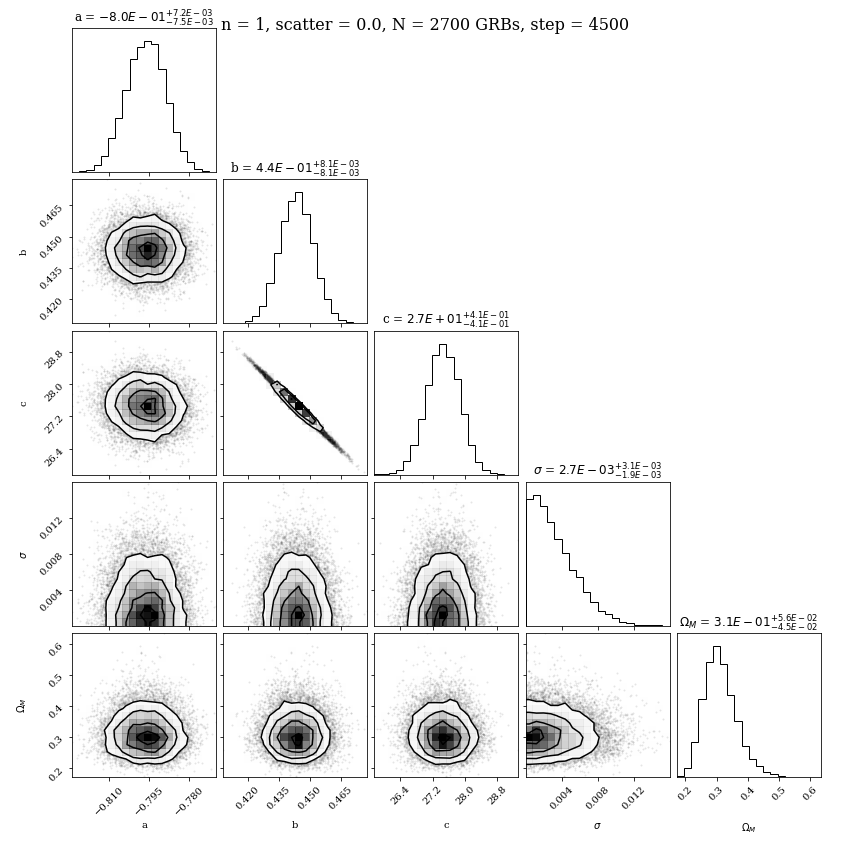}
    \includegraphics[scale=0.18]{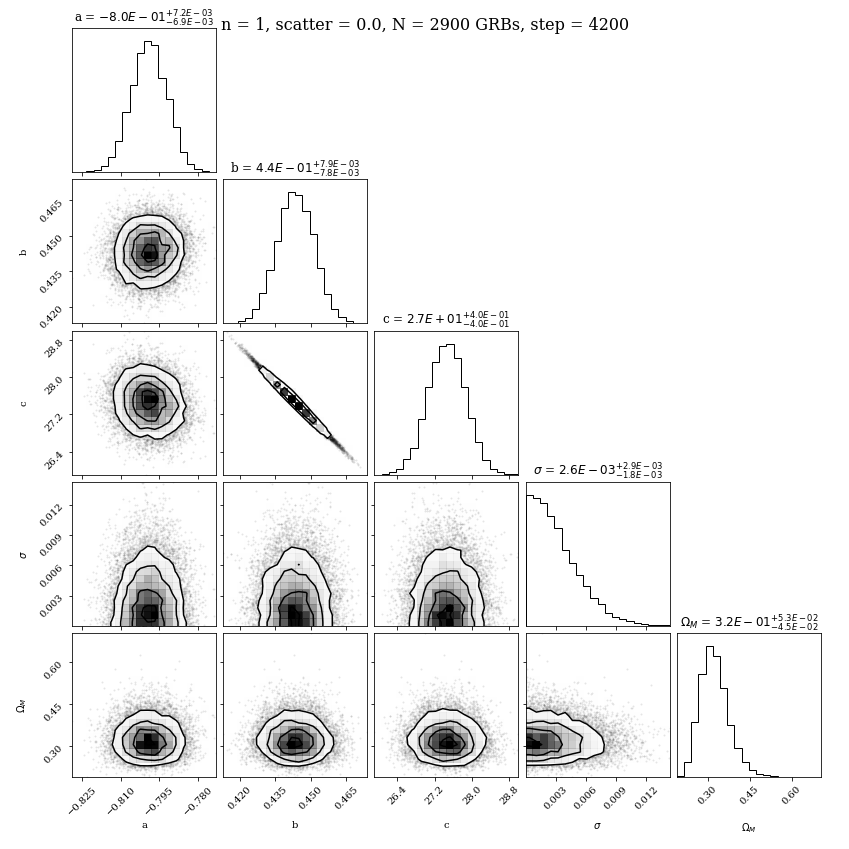}
    \caption{\textbf{Left panel.} Contour plots with the parameters of the fundamental plane relation in X-rays ($a,b,c$) together with the intrinsic scatter $\sigma$ and $\Omega_M$, in the case of 2700 simulated GRBs with the evolution correction through the Efron \& Petrosian method. \textbf{Right panel.} The same of the left panel, considering 2900 simulated GRBs.}
    \label{fig2}
\end{figure}

To both understand and predict the applicability of GRBs as standalone standard candles, we simulate additional GRBs to see how many are needed to reach uncertainties on $\Omega_M$ comparable to that of SNe Ia-derived values; we compare against Conley et al. (2011)\cite{Conley2010}, Betoule et al. (2014)\cite{Betoule2014}, and Scolnic et al. (2018)\cite{Scolnic2018} symmetrized error and standard deviation limits. In Figure \ref{fig2}, the contours for the parameters of the fundamental plane together with $\Omega_M$ after the Efron \& Petrosian method application are shown in the case of 2700 (left) and 2900 (right) simulated GRBs. To do so, we first use both the full optical and X-ray GRB samples as independent bases for GRB simulation, conducted again by MCMC techniques. We find that the optical sample yields much smaller uncertainties on each simulation than the X-ray, and consequently a more constrained value of $\Omega_M$. To increase this precision, we explore two methods of trimming our GRB sample leading to tighter planes that are in turn used as the base for simulation; an a posteriori trim using the smallest sample of GRBs for which the intrinsic scatter on the 3D plane which they defined remained near-zero, and an ‘a priori’ trim for which a number of possible sample sizes were tested and the one that yielded the best results after the fact is chosen. By simulating a large range of GRB sample sizes in both optical and X-ray wavelengths, we find by the construction of probability maps that the same precision as Betoule et al. (2014)\cite{Betoule2014} is achieved with only 376 simulated optical GRBs for which 47.5\% of the fundamental plane variable error bars have been halved by a light curve reconstruction procedure. We find that the Conley et al. (2011)\cite{Conley2010} limit is already achievable in most circumstances with current GRB numbers, and that the Scolnic et al. (2018)\cite{Scolnic2018} limit is achieved for 1152 optical GRBs. Considering both the detection power of future deep-space surveys THESEUS\cite{Stratta2018} and SVOM\cite{cordier2015svom,cordier2018svom} and the nearing capability of machine learning approaches to extract unknown GRB redshifts, we conclude that GRBs will be as efficient standalone probes as SNe Ia by the year 2038. These results are interesting because, as the definition of GRBs as standard candles becomes more reliable with the introduction of the optical and X-ray fundamental planes, the addition of these astrophysical objects to SNe Ia and BAO data will soon give the most precise derivation of $\Omega_M$ ever achieved.

\section{Summary and Conclusions}
In this work, it was highlighted how and to what extent the GRBs may contribute to the cosmological analysis in the future. In the first part (1), the issue of the Hubble constant tension has been investigated with the Pantheon sample of SNe Ia, showing how the $H_0$ itself is characterized by a slowly decreasing trend with the redshift. This opens the discussion on the reason for this result: together with the possibility that a modified gravity theory could be an explanation, it is also likely that different selection biases persist in the observations of SNe Ia. This strongly suggests that the methods for biases correction, such as the Efron \& Petrosian method, are needed in many astrophysical probes to achieve a reliable estimation of the cosmological parameters. Nevertheless, the limited coverage of redshift achieved by SNe Ia leads to the need for probes that can be observed at larger redshift. To this end, in the second part (2) it was shown how the 3D fundamental plane relation in the optical may help the future standardization of GRBs together with its counterpart in the X-rays. In addition, through the simulation of GRBs, it has been shown how this probe is a promising candidate to extend the Hubble diagram up to redshift greater than the ones of SNe Ia. Furthermore, a particular class of objects arouses much interest in the latest years, namely the class of LGRBs associated with Supernovae Ib/c (GRB-SNe): these manifest as supernovae appearing in the afterglow of the LGRBs and are very important since it was highlighted how the associated SNe obey a stretch-luminosity relation similar to the typical one of SNe Ia\cite{Cano2014}. The physics behind their emission mechanism has been an object of study on several occasions\cite{Dainotti2008GRBSN,Dainotti2010GRBSN,Cano2017} and one of the most interesting features of this class is that within the $L_X-T^*_X$ relation they show a Spearman correlation coefficient higher than the other subclasses of GRBs\cite{Dainotti2017a}. Thus, it is possible to use these events as a bridge between the properties of LGRBs and the ones of SNe Ib/c, giving a new perspective on the forthcoming standardization of GRB-SNe Ib/c. It is expected that the next observations of the new transients in the optical through the Subaru Telescope\cite{Watanabe2001} and KISO Telescope\cite{Sako2018} will help to investigate for selection biases and correct the current cosmological expectations on the evolution of the universe. Reliable testing of the cosmological models requires always new and more distant standardizable candles, and the GRBs have proven to be a reliable candidate for this purpose.

\bibliographystyle{ws-procs961x669}
\bibliography{main}

\end{document}